\documentclass[epj]{svjour}
\usepackage{graphics}
\begin{document}
\title{Recent developments in the constituent quark model 
including quark-antiquark pairs}
\author{Elena Santopinto\inst{1} \and Roelof Bijker\inst{2}
}                     
%
%
\institute{INFN, Sezione di Genova, via Dodecaneso 33, 
16164 Genova, Italy \and 
ICN-UNAM, AP 70-543, 04510 Mexico DF, Mexico}
\titlerunning{Constituent quark model including quark-antiquark pairs}
\date{Received: date / Revised version: date}
%
\abstract{
We present the formalism for a new generation of unquenched 
quark models for baryons in which the effects of quark-antiquark pairs 
($u \bar{u}$, $d \bar{d}$ and $s \bar{s}$) are taken into account in an 
explicit form via a microscopic, QCD-inspired, quark-antiquark creation 
mechanism. The present approach is an extension of the flux-tube breaking 
model of Geiger and Isgur in which now the contributions of quark-antiquark 
pairs can be studied for any inital baryon and for any flavor of the $q \bar{q}$ 
pair. It is shown that the inclusion of $q \bar{q}$ pairs leads to a large 
contribution of orbital angular momentum to the proton spin. 
\PACS{
      {14.20.Dh}{Protons and neutrons}   \and
      {12.39.-x}{Phenomenological quark models} \and
      {11.30.Hv}{Flavor symmetries} 
     } 
} 
\maketitle
\section{Introduction}

One of the main goals of hadronic physics is to understand the structure 
of the nucleon and its excited states in terms of effective degrees of freedom and, 
at a more fundamental level, the emergence of these effective degrees of freedom 
from QCD, the underlying theory of quarks and gluons \cite{Isgur}. 
Despite the progress made in lattice calculations, it remains a daunting problem 
to solve the QCD equations in the non-perturbative region. Therefore, one has to 
rely on effective models of hadrons, such as the constituent quark model (CQM).  

\begin{figure}[ht]
\centering
\resizebox{0.45\textwidth}{!}{\includegraphics{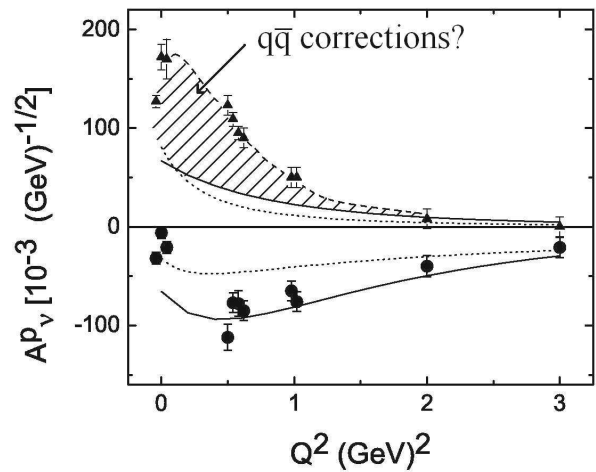}}
\caption[]{Transition form factors for the $D_{13}(1520)$ resonance. 
Experimental data are compared with theoretical predictions from the collective 
$U(7)$ model \cite{bil} (dotted line) and the hypercentral model \cite{hypercentral} 
(solid line).}
\label{d13}
\end{figure}

There exists a large variety of CQMs, among others 
the Isgur-Karl model \cite{IK}, the Capstick-Isgur model \cite{capstick}, 
the collective model \cite{bil}, the hypercentral model \cite{hypercentral}, 
the chiral boson-exchange model \cite{olof} and the Bonn instanton model 
\cite{bn}. While these models display important and peculiar differences, 
they share the main features: the effective degrees of freedom of three 
constituent quarks ($qqq$ configurations), the $SU(6)$ spin-flavor symmetry 
and a long-range confining potential. All of these models reproduce the mass 
spectrum of baryon resonances reasonably well. At the same time, they show 
very similar deviations 
for other properties, such as for example the photocouplings. Since the 
photocouplings depend mostly on the spin-flavor structure, all models that  
have the same $SU(6)$ structure in common, show the same behavior, {\it e.g.} the 
photocouplings for the $\Delta(1232)$ are underpredicted by a large amount, even 
though their ratio is reproduced correctly. In general, the helicity amplitudes  
(or transition form factors) show deviations from CQM calculations at low values of 
$Q^2$. As an illustration we show in Fig.~\ref{d13} the transverse electromagnetic 
transition form factors of the $D_{13}(1520)$ resonance for different CQMs. 
The problem of missing strength at low $Q^2$ can be attributed to the lack of 
explicit quark-antiquark degrees of freedom, which become more important in 
the outer region of the nucleon.

Additional evidence for higher Fock components in the baryon wave function 
($qqq-q\bar{q}$ configurations) comes from CQM studies of the 
electromagnetic and strong decay widths of $\Delta(1232)$ and $N(1440)$, 
the spin-orbit splitting of $\Lambda(1405)$ and $\Lambda(1520)$, the 
low $Q^2$ behavior of transition form factors, and the large $\eta$ decay 
widths of $N(1535)$, $\Lambda(1670)$ and $\Sigma(1750)$. More direct 
evidence for the importance of quark-antiquark components in the proton 
comes from measurements of the $\bar{d}/\bar{u}$ asymmetry in the nucleon sea 
\cite{Kumano,Garvey} and parity-violating electron scattering experiments, 
which report a nonvanishing strange quark contribution, albeit small, to 
the charge and magnetization distributions \cite{Acha}. 

The role of higher Fock components in the CQM has been studied in a series 
of papers by Riska {\em et al.} \cite{riska} in which it was shown that an 
appropriate admixture of the lowest $q^4 \bar{q}$ configurations may reduce 
the observed discrepancies between experiment and theory for several low-lying 
baryon resonances. The importance of mesonic contributions to the spin and 
flavor structure of the nucleon is reviewed in \cite{Kumano,Speth}. In the CQM 
based approach by Geiger and Isgur, the effects of quark-antiquark pairs 
were included in a flux-tube breaking model based on valence-quark plus glue 
dominance to which $q \bar{q}$ pairs are added in perturbation 
\cite{mesons,baryons}. 
The latter approach has the advantage that the effects of quark-antiquark 
pairs are introduced into the CQM via a QCD-inspired pair-creation mechanism, 
which opens the possibility to study the importance of effects of $q \bar{q}$ 
pairs in baryon structure in a systematic and unified way. 

The aim of the this contribution is to present a generalization of the flux-tube  
breaking model of \cite{baryons}. The resulting unquenched quark model 
is valid for any initial baryon (or baryon resonance) and for any flavor of the 
quark-antiquark pair (not only $s \bar{s}$ as in \cite{baryons}, but also 
$u \bar{u}$ and $d \bar{d}$), and can be applied to any model of baryons and mesons. 
As a test of the formalism, we present some results in the closure limit. 
Finally, we discuss an application of the unquenched quark model 
to the spin of the proton. In a separate contribution, we present an application 
to the flavor asymmetry of the nucleon sea \cite{roelof}. 

\section{Unquenched quark model}

In the flux-tube model for hadrons, the quark potential model arises from an 
adiabatic approximation to the gluonic degrees of freedom embodied in the flux 
tube \cite{flux}. The role of quark-antiquark pairs in meson spectroscopy was 
studied in a flux-tube breaking model \cite{mesons} in which 
the $q \bar{q}$ pair is created with the $^{3}P_0$ quantum numbers of the vacuum.  
Subsequently, it was shown by Geiger and Isgur \cite{OZI} that a {\it miraculous}  
set of cancellations between apparently uncorrelated sets of intermediate states 
occurs in such a way that they compensate each other and do not destroy the good 
CQM results for the mesons. In particular, the OZI hierarchy is preserved and 
there is a near immunity of the long-range confining potential, since the change 
in the linear potential due to the creation of quark-antiquark pairs in the string 
can be reabsorbed into a new strength of the linear potential, {\em i.e.} in a new 
string tension. As a result, the net effect of the mass shifts from pair creation 
is smaller than the naive expectation of the order of the strong decay widths. 
However, it is necessary to sum over  large towers of intermediate 
states to see that the spectrum of the mesons, after unquenching and renormalizing, 
is only weakly perturbed. An important conclusion is that no simple truncation of 
the set of meson loops is able to reproduce such results \cite{OZI}.

\begin{figure}
\centering
\resizebox{0.45\textwidth}{!}{\includegraphics{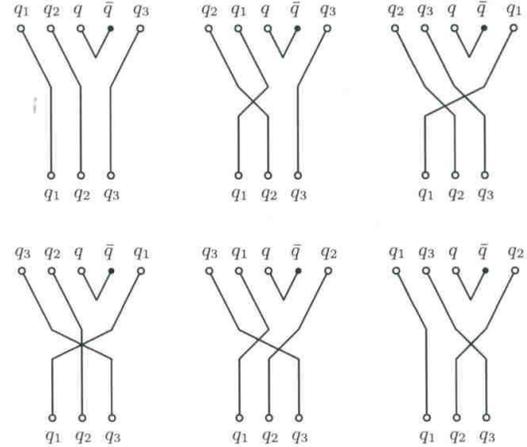}}
\caption{Quark line diagrams for $A \rightarrow B C$ with 
$q_1 q_2 q_3 = uud$ and $q \bar{q} = s \bar{s}$}
\label{diagrams}
\end{figure}

The extension of the flux-tube breaking model to baryons requires a proper treatment 
of the permutation symmetry between identical quarks. As a first step, Geiger and Isgur 
investigated the importance of $s \bar{s}$ loops in the proton by taking into account 
the contribution of the six different diagrams of Fig.~\ref{diagrams} with 
$q \bar{q}=s \bar{s}$ and $q_1 q_2 q_3 = uud$, and by using 
harmonic oscillator wave functions for the baryons and mesons \cite{baryons}. 
In the conclusions, the authors emphasized: {\em It also seems very worthwhile 
to extend this calculation to $u \bar{u}$ and $d \bar{d}$ loops. Such an extension 
could reveal the origin of the observed violations of the Gottfried sum rule and 
also complete our understanding of the origin of the spin crisis.} 
In this contribution, we take up this challenge and present a generalization 
of the formalism of \cite{baryons} in which quark-antiquark contributions 
can be studied 
\begin{itemize}
\item for any initial baryon resonance, 
\item for any flavor of the quark-antiquark pair, and 
\item for any model of baryons and mesons.
\end{itemize}
These extensions were made possible by two developments: the solution of the 
problem of the permutation symmetry between identical quarks by means of 
group-theoretical techniques, and the construction of an algorithm to 
generate a complete set of intermediate states for any model of baryons 
and mesons. While the first improvement allows the evaluation of the 
contribution of quark-antiquark pairs for any initial baryon $q_1 q_2 q_3$ 
(ground state or resonance) and for any flavor of the $q \bar{q}$ pair 
(not only $s\bar{s}$, but also $u\bar{u}$ and $d\bar{d}$), the second one 
permits the carry out the sum over intermediate states up to saturation for 
any model of baryons and mesons, as long as their wave functions are expressed 
in the basis of harmonic oscillator wave functions. 

The ensuing unquenched quark model is based on an adiabatic treatment of the 
flux-tube dynamics to which $q \bar{q}$ pairs with vacuum quantum numbers are 
added as a perturbation \cite{baryons}. The pair-creation mechanism is 
inserted at the quark level and the one-loop diagrams are calculated by 
summing over a complete set of intermediate states. Under these assumptions, 
to leading order in pair creation, the baryon wave function is given by 
\begin{eqnarray} 
\mid \psi_A \rangle &=& {\cal N} \left[ \mid A \rangle 
+ \sum_{BC l J} \int d \vec{k} \, \mid BC \vec{k} \, l J \rangle \right. 
\nonumber\\
&& \hspace{2cm} \times \left. 
\frac{ \langle BC \vec{k} \, l J \mid T^{\dagger} \mid A \rangle } 
{M_A - E_B - E_C} \right] ~.
\end{eqnarray}
Here $T^{\dagger}$ is the $^{3}P_0$ quark-antiquark pair-creation operator 
\cite{roberts}, $A$ denotes the initial baryon and $B$ and $C$ the intermediate 
baryons and meson, $\vec{k}$ and $l$ represent the relative radial momentum and 
orbital angular momentum of $B$ and $C$, and $J$ is the total angular momentum 
$\vec{J} = \vec{J}_B + \vec{J}_C + \vec{l}$. 

In general, matrix elements of an observable $\hat{\cal O}$ can be expressed as 
\begin{eqnarray}
{\cal O} = \langle \psi_A \mid \hat{\cal O} \mid \psi_A \rangle 
= {\cal O}_{\rm val} + {\cal O}_{\rm sea} ~,
\end{eqnarray}
where the first term denotes the contribution from the valence quarks  
\begin{eqnarray}
{\cal O}_{\rm val} &=& {\cal N}^2 \langle A \mid \hat{\cal O} \mid A \rangle 
\end{eqnarray}
and the second term that from the $q \bar{q}$ pairs
\begin{eqnarray}
{\cal O}_{\rm sea} &=& {\cal N}^2 \sum_{BC l J} \int d \vec{k} \,
\sum_{B'C' l' J'} \int d \vec{k}^{\, \prime} \,
\frac{ \langle A \mid T \mid B' C' \vec{k}^{\, \prime} \, l' J' \rangle } 
{M_A - E_{B'} - E_{C'}} 
\nonumber\\
&& \langle B' C' \vec{k}^{\, \prime} \, l' J' \mid \hat{\cal O}  
\mid B C \vec{k} \, l J \rangle \, 
\frac{ \langle B C \vec{k} \, l J \mid T^{\dagger} \mid A \rangle } 
{M_A - E_B - E_C} ~.
\label{me}
\end{eqnarray}
As mentioned before, we developed an algorithm based upon group-theoretical 
techniques to generate a complete set of intermediate states of good 
permutational symmetry, which makes it possible to perform the sum over 
intermediate states up to saturation, and not just for the first few shells 
as in \cite{baryons}. Not only does this have a significant impact on the 
numerical result, but it is necessary for consistency 
with the OZI-rule and the success of CQMs in hadron spectroscopy. 

\section{Closure limit}

The evaluation of the contribution of the quark-antiquark pairs simplifies 
considerably in the closure limit, which arises when the energy denominators 
in Eq.~(\ref{me}) do not depend on the quantum numbers of the intermediate states. 
In this case, the sum over the complete set of intermediate states can be solved 
by closure and the contribution of the quark-antiquark pairs to the matrix element 
reduces to 
\begin{eqnarray}
{\cal O}_{\rm sea} &\propto& \langle A \mid T \, \hat{\cal O} \,   
T^{\dagger} \mid A \rangle ~.
\end{eqnarray}
Moreover, if the pair-creation operator does not couple to the motion of the 
valence quarks, the valence quarks act as spectators. In this case, the contribution 
of the $q \bar{q}$ pairs reduces to the expectation value of ${\cal O}$ between the 
$^{3}P_0$ pair states created by $T^{\dagger}$ 
\begin{eqnarray}
{\cal O}_{\rm sea} &\propto& \langle 0 \mid T \, \hat{\cal O} \,   
T^{\dagger} \mid 0 \rangle ~, 
\end{eqnarray}
the so-called closure-spectator limit. 
Especially when combined with symmetries, the closure limit not only provides simple 
expressions for the relative flavor content of physical observables, but also can give 
further insight into the origin of cancellations between the contributions from different 
intermediate states. 

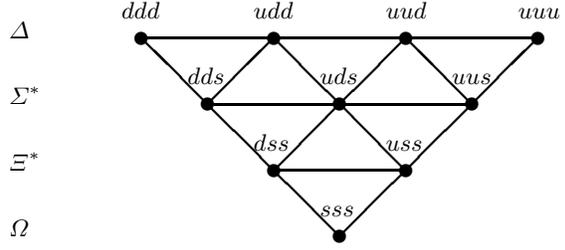
\begin{figure}[t]
\centering
\setlength{\unitlength}{0.5pt}
\begin{picture}(440,200)(20,80)
\thicklines
\put(100,250) {\line(1,0){300}}
\put(150,200) {\line(1,0){200}}
\put(200,150) {\line(1,0){100}}

\put(150,200) {\line(1,1){ 50}}
\put(200,150) {\line(1,1){100}}
\put(250,100) {\line(1,1){150}}

\put(250,100) {\line(-1,1){150}}
\put(300,150) {\line(-1,1){100}}
\put(350,200) {\line(-1,1){ 50}}

\multiput(100,250)(100,0){4}{\circle*{10}}
\multiput(150,200)(100,0){3}{\circle*{10}}
\multiput(200,150)(100,0){2}{\circle*{10}}
\put(250,100){\circle*{10}}

\put( 85,265){$ddd$}
\put(185,265){$udd$}
\put(285,265){$uud$}
\put(385,265){$uuu$}
\put(135,215){$dds$}
\put(235,215){$uds$}
\put(335,215){$uus$}
\put(185,165){$dss$}
\put(285,165){$uss$}
\put(235,115){$sss$}

\put(0,250){$\Delta$}
\put(0,200){$\Sigma^{\ast}$}
\put(0,150){$\Xi^{\ast}$}
\put(0,100){$\Omega$}

\end{picture}
\caption[]{\small Ground state decuplet baryons}
\label{decuplet}
\end{figure}

\begin{table}[b]
\centering
\caption[]{\small Relative contributions of $\Delta u$, $\Delta d$ and $\Delta s$ 
in the closure limit to the spin of the ground state decuplet baryons}   
\label{spindec}
\begin{tabular}{ccccccc}
\hline
& & & & & & \\
$qqq$ & $^{4}10[56,0^+]$ & $\Delta u$ &:& $\Delta d$ &:& $\Delta s$ \\
& & & & & & \\
\hline
& & & & & & \\
$uuu$ & $\Delta^{++}$ & 9 &:& 0 &:& 0 \\
$uud$ & $\Delta^{+}$ & 6 &:& 3 &:& 0 \\
$udd$ & $\Delta^{0}$ & 3 &:& 6 &:& 0 \\
$ddd$ & $\Delta^{-}$ & 0 &:& 9 &:& 0 \\
$uus$ & $\Sigma^{\ast \, +}$ & 6 &:& 0 &:& 3 \\
$uds$ & $\Sigma^{\ast \, 0}$ & 3 &:& 3 &:& 3 \\
$dds$ & $\Sigma^{\ast \, -}$ & 0 &:& 6 &:& 3 \\
$uss$ & $\Xi^{\ast \, 0}$ & 3 &:& 0 &:& 6 \\
$dss$ & $\Xi^{\ast \, -}$ & 0 &:& 3 &:& 6 \\
$sss$ & $\Omega^{-}$ & 0 &:& 0 &:& 9 \\
& & & & & & \\
\hline
\end{tabular}
\end{table}

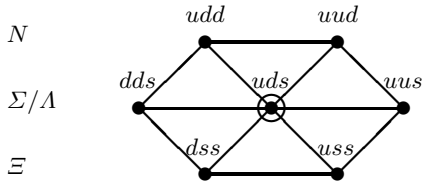
\begin{figure}[t]
\centering
\setlength{\unitlength}{0.5pt}
\begin{picture}(340,150)(50,130)
\thicklines
\put(200,250) {\line(1,0){100}}
\put(150,200) {\line(1,0){200}}
\put(200,150) {\line(1,0){100}}

\put(150,200) {\line(1,1){ 50}}
\put(200,150) {\line(1,1){100}}
\put(300,150) {\line(1,1){ 50}}

\put(200,150) {\line(-1,1){ 50}}
\put(300,150) {\line(-1,1){100}}
\put(350,200) {\line(-1,1){ 50}}

\multiput(200,250)(100,0){2}{\circle*{10}}
\multiput(150,200)(100,0){3}{\circle*{10}}
\multiput(200,150)(100,0){2}{\circle*{10}}
\put(250,200){\circle{20}}

\put(185,265){$udd$}
\put(285,265){$uud$}
\put(135,215){$dds$}
\put(235,215){$uds$}
\put(335,215){$uus$}
\put(185,165){$dss$}
\put(285,165){$uss$}

\put(50,250){$N$}
\put(50,200){$\Sigma/\Lambda$}
\put(50,150){$\Xi$}

\end{picture}
\caption[]{\small Ground state octet baryons}
\label{octet}
\end{figure}

As an example, we discuss some results for the operator 
\begin{eqnarray}
\Delta q = 2 \left< S_z(q) + S_z(\bar{q}) \right> ~,
\end{eqnarray}
which determines the fraction of the baryon's spin carried by the quarks 
and antiquarks with flavor $u$, $d$ and $s$. First, we consider the ground state 
decuplet baryons with $^{4}10 [56,0^+]_{3/2}$ of Fig.~\ref{decuplet}. 
Since the valence-quark configuration of the $\Delta$ resonances does not contain 
strange quarks, the contribution $\Delta s$ of the $s \bar{s}$ pairs to the spin 
is given by the closure-spectator limit which vanishes due to the properties of the 
$^{3}P_0$ wave functions. The same holds for the contribution 
of $d \bar{d}$ pairs to the $\Delta^{++}$, $\Sigma^{\ast \, +}$, $\Xi^{\ast \, 0}$ 
and $\Omega^{-}$ resonances, and that of $u \bar{u}$ pairs to the $\Delta^{-}$, 
$\Sigma^{\ast \, -}$, $\Xi^{\ast \, -}$ and $\Omega^{-}$ resonances. 
In the closure limit the relative contribution of the quark flavors from the 
quark-antiquark pairs to the baryon spin is the same as that from the valence quarks
\begin{eqnarray}
\Delta u_{\rm sea} : \Delta d_{\rm sea} : \Delta s_{\rm sea} = 
\Delta u_{\rm val} : \Delta d_{\rm val} : \Delta s_{\rm val} ~.
\end{eqnarray}
The latter property is a consequence of the spin-flavor symmetry of the ground state 
baryons and holds for both the decuplet with quantum numbers $^{4}10 [56,0^+]_{3/2}$ 
and the octet with $^{2}10 [56,0^+]_{1/2}$ (see Fig.~\ref{octet}). 
Table~\ref{spindec} shows the relative contributions of $\Delta u$, $\Delta d$ 
and $\Delta s$ to the spin of the ground state decuplet baryons in the closure limit. 
The results for the ground state octet baryons are given in Table~\ref{spinoct}. 
Finally, the results for the decuplet and octet baryons are related by
\begin{eqnarray}
\left( \Delta u + \Delta d + \Delta s \right)_{\rm decuplet} = 
3\left( \Delta u + \Delta d + \Delta s \right)_{\rm octet} ~.
\end{eqnarray}

\begin{table}[b]
\centering
\caption[]{\small Relative contributions of $\Delta u$, $\Delta d$ and $\Delta s$ 
in the closure limit to the spin of the ground state octet baryons}   
\label{spinoct}
\begin{tabular}{ccrcrcr}
\hline
& & & & & & \\
$qqq$ & $^{2}8[56,0^+]$  & $\Delta u$ &:& $\Delta d$ &:& $\Delta s$ \\
& & & & & & \\
\hline
& & & & & & \\
$uud$ & $p$ & 4 &:& $-1$ &:& 0 \\
$udd$ & $n$ & $-1$ &:& 4 &:& 0 \\
$uus$ & $\Sigma^+$ & 4 &:& 0 &:& $-1$ \\
$uds$ & $\Sigma^0$ & 2 &:& 2 &:& $-1$ \\
      & $\Lambda$  & 0 &:& 0 &:& 3 \\
$dds$ & $\Sigma^-$ & 0 &:& 4 &:& $-1$ \\
$uss$ & $\Xi^0$ & $-1$ &:& 0 &:& 4 \\
$dss$ & $\Xi^-$ & 0 &:& $-1$ &:& 4 \\
& & & & & & \\
\hline
\end{tabular}
\end{table}

At a qualitative level, a vanishing closure limit helps to explain the 
phenomenological success of CQMs. As an example, the strange content of 
the proton which vanishes in the closure limit, is expected to be small, 
in agreement with the experimental data from parity-violating electron 
scattering (for the most recent data see \cite{Acha}). 

In addition, the results in Tables~\ref{spindec} and \ref{spinoct} impose 
very stringent conditions on the numerical calculations, since each entry 
involves the sum over a complete set of intermediate states. Therefore, the 
closure limit provides a highly nontrivial test of the computer codes which 
involves both the spin-flavor sector, the permutation symmetry, the 
construction of a complete set of intermediate states and the implementation 
of the sum over all of these states. 

\section{Proton spin} 

The unquenched quark model makes it possible to study the effect of 
quark-antiquark pairs on the fraction of the proton spin carried by quarks.  
Ever since the European Muon Collaboration at CERN 
showed that {\it the total quark spin constitutes a rather small fraction of 
the spin of the nucleon} \cite{emc}, there has 
been an enormous interest in the spin structure of the proton \cite{protonspin}. 
The most recent value for the contribution of the quark spins is $33.0 \pm 3.9$ \% 
\cite{hermes1}. The total spin of the proton is distributed among valence and  
sea quarks, orbital angular momentum and gluons 
\begin{eqnarray}
\frac{1}{2} = \frac{1}{2} \left( \Delta u + \Delta d + \Delta s \right) + \Delta L 
+ \Delta G ~,
\end{eqnarray}
where
\begin{eqnarray}
\Delta q = \int_0^1 dx \left[ q_{\uparrow}(x) + \bar{q}_{\uparrow}(x) 
-q_{\downarrow}(x) - \bar{q}_{\downarrow}(x) \right] 
\end{eqnarray}
is the fraction of the proton's spin carried by the light quarks and antiquarks 
with flavor $q=u$, $d$, $s$. $\Delta L$ and $\Delta G$ represent the contributions from 
orbital angular momentum and gluons, respectively. There is increasing evidence 
that the gluon contribution is small (either positive or negative) and compatible 
with zero \cite{gluonexp,gluonth}, which indicates that the missing spin of the 
proton must be attributed to orbital angular momentum of the quarks and antiquarks. 

In the unquenched quark model, there are no explicit gluons, so the last term 
$\Delta G$ is absent from the outset. The fraction of the spin of the proton 
carried by the quarks is obtained from  
\begin{eqnarray}
\Delta q = 2 \left< S_z(q) + S_z(\bar{q}) \right> ~.
\end{eqnarray}
We carried out a calculation in which the parameters were taken from the literature 
\cite{baryons,CR}, and found that a large part of the proton spin is due to orbital 
angular momentum. More specifically, the $q \bar{q}$ pairs contribute about half 
of the proton spin, of which one quarter is due to the spin of the sea quarks and 
three quarters to orbital angular momentum. Similar conclusions regarding the 
importance of the contribution of orbital angular momentum to the proton spin were 
reached in studies with meson-cloud models \cite{cbm} and with axial exchange 
currents \cite{alfons}.   

\section{Summary, conclusions and outlook}

In this contribution, we have discussed the importance of quark-antiquark pairs 
in baryon spectroscopy. We have proposed a method based on the flux-tube 
breaking model that was originally introduced by Kokoski and Isgur for mesons 
\cite{mesons} and later extended by Geiger and Isgur to study the effects 
of $s \bar{s}$ loops in the proton \cite{baryons}. Here, we have presented a new 
generation of unquenched quark models for baryons by including, in addition 
to $s \bar{s}$ loops, the contributions of $u \bar{u}$ and $d \bar{d}$ loops 
as well.  

As a first application, we have applied the closure limit of the model - in 
which all intermediate states are degenerate - to the flavor decomposition 
of the spin of the ground state octet and decuplet baryons. It has been found that 
the relative contributions of the quark flavors from the $q \bar{q}$ pairs are 
the same as that of the valence quarks. The closure limit not only provides 
a stringent test of the numerical results, but also sheds light on the 
physical properties of (unquenched) quark models. 

Next, the unquenched quark model has been applied to the problem of the spin crisis 
of the proton. It was shown that the inclusion of $q \bar{q}$ pairs leads to a 
relatively large contribution (about 40 \%) of orbital angular momentum to the 
proton spin. In a separate contribution, we discuss an application to the flavor 
asymmetry of the nucleon sea \cite{roelof}.  

The present formalism is, obviously within the assumptions of the approach, 
valid for any initial baryon, any flavor of the $q \bar{q}$ pairs and any model 
of hadron structure. As such, it holds great promise in its ability to address 
in a general and systematic way a large number of open problems in the structure 
and spectroscopy of light baryons such as, among others, the flavor asymmetry of 
the nucleon sea, the spin crisis of the proton, the electromagnetic and strong 
couplings, the electromagnetic elastic and transition form factors of baryon 
resonances, their sea quark content and their flavor decomposition \cite{BS}. 

\section*{Acknowledgments}

This work was supported in part by a grant from INFN, Italy 
and in part by CONACYT, Mexico.

\end{document}